# Using Cloning to Solve NP Complete Problems

John A. Drakopoulos  and  Theodore N. Tomaras

Department of Physics, University of Crete
P.O.Box 2208, Heraklion 71003, Crete, Greece.
Tel. +3081-394246,   Fax +3081-394201
http://www.physics.uch.gr/~jad
e-mail: jad@physics.uch.gr

*ABSTRACT*
*Assuming a cloning oracle, satisfiability, which is an NP complete problem, is shown to belong to $BPP^C$ and $BQP^C$ (depending on the ability of the oracle **C** to clone either a binary random variable or a qubit). The same result is extended in the case of an approximate cloning oracle, thus establishing that $NP \subseteq BPP^C \subseteq BQP^C$ and $NP \subseteq BPP^{AC} \subseteq BQP^{AC}$, where **C** and AC are the exact and approximate cloning oracles respectively. Although exact cloning is impossible in quantum systems, approximate cloning remains a possibility. However, the best known methods for approximate cloning (based on unitary evolution) do not currently achieve the desired precision levels.  And it remains an open question whether they could be improved when non-linear (or non-unitary) operators are used. Finally, a straightforward attempt to dispense with cloning, replacing it by unitary evolution, is proved to be impossible.*

## 1. Introduction

Quantum computing is a new and exciting interdisciplinary area that combines computer science and (quantum) physics. As early as 1982, Feynman observed that the straightforward simulation of a quantum system on a classical computer (deterministic or probabilistic Turing machine) required an exponential slowdown and without any apparent way to speed up the simulation [11, 12]. He asked whether that is inherent to quantum systems and he suggested the design of computing machines based on quantum theory implying, at the same time, that such quantum computers could perhaps compute more efficiently than classical computers.

About the same time, and addressing the opposite problem, Benioff showed that a deterministic Turing machine could be simulated by the unitary evolution of a quantum process and thus provided the first indication of the strength of quantum computing [1, 2].

Subsequently, Deutsch proposed a general model of quantum computation--the *quantum Turing machine*--which could simulate any given quantum system but possibly with exponential slowdown [8].

Berstein and Vazirani improved upon the concept of a quantum Turing machine proposing a *univesal quantum Turing machine,* which, as they proved, could simulate a broad class of quantum Turing machines with only a polynomial slowdown [4].

In classical computing, logic circuits provide an alternative to Turing machines (actually current computers are build from integrated circuits and not as Turing machines). Deutsch first proposed a similar model of quantum circuits, which he called *quantum computational networks,* and examined some of their properties [9]. Yao subsequently proved the polynomial equivalence between quantum Turing machines and quantum circuits (thus providing a truly universal model of quantum computation) by proving that an arbitrary quantum Turing machine could simulate, and be simulated by, a polynomial size quantum circuit [20].

The model of quantum computing having been defined, its computational power must be identified. The following classes of decision problems have been defined:

P is the class of all decision problems that can be solved in polynomial time by a deterministic Turing machine

BPP is the class of all decision problems that can be solved in polynomial time by a probabilistic Turing machine with a probability of error bounded by 1/3 for all inputs

NP is the class of all decision problems that can be solved in polynomial time by a non-deterministic Turing machine

BQP is the class of all decision problems that can be solved in polynomial time by a quantum Turing machine with a probability of error bounded by 1/3 for all inputs

(Apparently it is P $\subseteq$ BPP $\subseteq$ NP. Furthermore, Benniof's result [1, 2] implies that P $\subseteq$ BQP.)

Deutsch and Josza [11] and Berthiaume and Brassard [10, 6] proved that, relative to certain oracles, there are computational problems that can be solved exactly and in polynomial time by quantum Turing machines but cannot be solved polynomially for all inputs by deterministic or probabilistic Turing machines. However, those problems belong to BPP; and thus the above results do not confer the supposed extra computing power of quantum Turing machines.

Bennett et al. [3] proved that relative to a random oracle, it is not true that NP $\subseteq$ BQP. Bernstein and Vazirani [4] proved that BPP $\subseteq$ BQP. Bernstein and Vazirani [4] and Simon [18] invented problems that are not known to be in BPP but belong to BQP. Shor gave polynomial time quantum algorithms for the factoring and discrete log problems [17]. (Note that not only Simon's problem but also factoring and discrete log belong to NP $\cap$ co-NP. However, Bennett et al. [3] proved that relative to a permutation oracle, it is not true that NP $\cap$ co-NP $\subseteq$ BQP.) Finally Grover showed how to accept the class NP relative to any oracle in time $O(2^{n/2})$. (A formal analysis of Grover's algorithm appears in [7].)

It should be noted that Shor's factoring and discrete log algorithms (with its implications in cryptography and cryptosystems) and Grover's database search algorithm are of practical importance. Furthermore, the application of quantum computing for the development of secure cryptographic communication systems (which detect unauthorized access or guarantee that information would not be compromised [15]) is obviously of high commercial value.

It is still unknown whether BPP $\subset$ BQP, and whether NP $\subseteq$ BQP. (The latter is tremendously important since the class NP contains a large number of optimization problems with a broad range of applications from computer science and statistics to engineering, automatic control, integrated circuit design etc.)

In this paper, we provide polynomial algorithms for solving the satisfiability problem, which is a known NP complete problem, assuming an oracle that can (either exactly or approximately) clone a binary random variable or a qubit. Therefore, we prove that NP $\subseteq$ BPP$^C$ $\subseteq$ BQP$^C$ and NP $\subseteq$ BPP$^{AC}$ $\subseteq$ BQP$^{AC}$, where **C**( ) and AC( ) are the exact and approximate cloning oracles as defined below. Such oracles, which do not merely provide a Boolean reply to a query but generate an object (such as a random variable or qubit), are called *manufacturing oracles*.

In section 2, we describe the satisfiability problem. In section 3, we provide an algorithm for the case of exact cloning, while, in section 4, we prove that the algorithm belongs to BPP$^C$ (or $\subseteq$ BQP$^C$ respectively). A similar construction is made for the case of approximate cloning in section 5. In section 6, an attempt is made to dispense with cloning and a negative result is shown. Finally, in concluding section 7, the limitations and some practical implications are discussed.

## 2. The satisfiability problem

A *Boolean variable x* is a variable that can assume only the values *true* and *false*, (which are usually associated with the numbers 1 and 0 respectively). A *truth assignment* is an assignment of true/false values to a set of Boolean variables.

A logical *expression* or *Boolean formula* is an expression consisting of Boolean variables combined with *logical connectives* such as *or*, *and*, *not*, etc. A logical expression can be *unsatisfiable* (when it is false for all truth assignments of its variables), *satisfiable* (when it is true for at least one assignment of its variables), and *tautological* (when it is true for all truth assignments of its variables).

For example, assume that + denotes logical disjunction (or), · denotes logical conjunction (and), and a bar over an expression denotes the logical negation of the expression. Then $(\overline{x_2} + \overline{x_3}) \cdot (x_1 + x_2 + x_4)$ is a logical expression containing four variables and is true if and only if at least one of $x_2$ and $x_3$ is false and at least one of $x_1$, $x_2$, and $x_4$ is true. Therefore, it is satisfiable but not tautological.

A *literal* is a Boolean variable or its negation, and a *clause* is a disjunction of literals. The *satisfiability* problem is as follows:

> Given *m* clauses $c_0, \ldots, c_{m-1}$ containing *n* Boolean variables $x_0, \ldots, x_{n-1}$ is the formula $c_0 \cdot c_1 \cdot \ldots \cdot c_{m-1}$ satisfiable?

Satisfiability is known to be NP-complete [16, ch.8, theorem 8.2].

Finally, if *E* is a logical expression and *S* a truth assignment *E/S* is the logical expression that results after we substitute the variables of *E* that appear in *S* with their assigned values. For example, if $E = a + b + c \cdot d$ and $S = \{a=false, d=true\}$ then $E/S = b + c$.

## 3. The logical boosting algorithm

We assume that [ ] is the *Boolean-to-integer* operator (that is $[\varphi] = 1$, if $\varphi$ is true; $[\varphi] = 0$, if $\varphi$ is false). We denote *probability measure* by $P(\ )$, and *quantum and, or, and not* gates by &, ||, and ~ respectively.

A *binary random variable* is a random variable that can be either 0 or 1. A *qubit*, which is the corresponding unit of quantum information, behaves more or less like a random variable and is described by a pair of complex numbers (*a*, *b*) whose magnitudes are the probabilities that the qubit would be in state 0 or state 1 (respectively). Since those are different concepts, for the rest of this paper, we shall use both the term binary random variable and qubit to describe situations where either object could be used. The subsequent results would thus apply to probabilistic and quantum computing (respectively).

The following algorithm assumes a function **C**(*X*) that can generate a copy (or clone) of any binary random variable or qubit *X*. In other words, **C**(*X*) returns a binary random variable or qubit that is independent of *X* and identically distributed.

---

**The logical boosting (LB(N)) algorithm:**

Given a binary random variable or qubit copy function **C**( ), a satisfiability problem of *m* clauses $c_0, \ldots, c_{m-1}$ and *n* Boolean variables $x_0, \ldots, x_{n-1}$ the algorithm decides whether $c_0 \cdot c_1 \cdot \ldots \cdot c_{m-1}$ is satisfiable as follows:

1. Create *n* independent binary random variables or qubits $X_0, \ldots, X_{n-1}$ (corresponding to Boolean variables $x_0, \ldots, x_{n-1}$) such that
   $$P(X_i = 1) = P(X_i = 0) = 1/2, \quad \text{for } i = 0,1,\ldots,n-1.$$
2. Create *m* Boolean random variables or qubits $C_0, \ldots, C_{m-1}$ that correspond to clauses $c_0, \ldots, c_{m-1}$:

   If $x_{i_0}, \ldots, x_{i_a}$ are the Boolean variables appearing non-negated in $c_j$ and

   $x_{k_0}, \ldots, x_{k_b}$ are the Boolean variables appearing negated in $c_j$

   then
   $$C_j = X_{i_0} \,\|\, \ldots \,\|\, X_{i_a} \,\|\, (\sim X_{k_0}) \,\|\, \ldots \,\|\, (\sim X_{k_b}), \quad j = 0, 1, \ldots, m-1.$$

3. Create Boolean random variable or qubits $D_0$ as follows:
   $$D_0 = C_0 \,\&\, \ldots \,\&\, C_{m-1}$$
4. Choose $N \geq n$ and create *N* Boolean random variables or qubits $D_1, \ldots, D_N$ as follows:
   $$D_v = D_{v-1} \,\|\, \mathbf{C}(D_{v-1}), \quad v = 1, 2, \ldots, N.$$
5. Look at $D_N$. If it is 1, the initial problem is satisfiable; otherwise assume that the problem is unsatisfiable.

---

Note that the algorithm requires &-gates and ||-gates with large number of inputs (*n* could be very large and *m* could be even larger). However, this situation can be easily remedied since an *M*-input &-gate or ||-gate can be constructed as a network of $\left\lceil \dfrac{M-1}{K-1} \right\rceil$ *K*-input &-gates or ||-gates respectively.

## 4. The probability of error and the time complexity of the LB algorithm

***Theorem 1.*** *The LB(N) algorithm asserts correctly that a given satisfiability problem S is satisfiable, and the probability of error $P_{err}$, in the case it asserts that the problem is unsatisfiable, is bounded:*

$$P_{err} < \left(\frac{1}{e}\right)^{2^{N-n}}$$

*where e is the Euler number (e ≅ 2.71828182845…), n is the number of Boolean variables, and N is the boosting level chosen in the LB algorithm.*

***Proof***
Let
       *u* be any truth assignment of the Boolean variables $x_0, \ldots, x_{n-1}$
       *u'* be the corresponding assignment of values 0 and 1 to qubits $X_0, \ldots, X_{n-1}$
       $k_S$ be the number of different truth assignments that satisfy *S*; that is,

$$k_S = \sum_u [S/u]$$
$$d_v = P(D_v = 0), \quad v = 0, 1, 2, \ldots, N$$

Then,
$$d_0 = 1 - P(D_0 = 1) = 1 - \sum_{u'} P(D_0 = 1/u') P(u') = 1 - \frac{1}{2^n} \sum_u [S/u] = 1 - \frac{k_S}{2^n}$$

and since $D_v$ and $\mathbf{C}(D_v)$ are independent and identically distributed:
$$d_{v+1} = P(D_v \| \mathbf{C}(D_v) = 0) = P(D_v = 0) P(\mathbf{C}(D_v) = 0) = d_v^2, \quad v = 0, 1, \ldots, N-1$$

As a result,
$$d_N = d_0^{2^N} = \left(1 - \frac{k_S}{2^n}\right)^{2^N} = \left(\left(1 - \frac{k_S}{2^n}\right)^{2^n}\right)^{2^{N-n}} < \left(\frac{1}{e^{k_S}}\right)^{2^{N-n}}$$

The above inequality emerges from the fact that the sequence $a_m = \left(1 - \frac{k_S}{m}\right)^m$, $m > 0$, is strictly increasing and converges to its supremum $e^{-k_S}$.

Now, observe that if $D_N$ is found to be 1, then
$$P(D_N = 1) > 0 \Rightarrow d_N < 1 \Rightarrow d_0 < 1 \Rightarrow P(D_0 = 1) > 0 \Rightarrow k_S > 0 \Rightarrow S \text{ is satisfiable.}$$

Thus, the algorithm finds the correct answer in this case. On the other hand, when $D_N$ is found to be 0 and $S$ is unsatisfiable, the algorithm again guesses correctly. The case of error is only when $D_N$ is found to be 0 and $S$ is satisfiable. However, in that case it is $k_S > 0$ (or equivalently $k_S \geq 1$) and thus the probability of error is

$$P_{err} = P(D_N = 0) = d_N < \left(\frac{1}{e^{k_S}}\right)^{2^{N-n}} \leq \left(\frac{1}{e}\right)^{2^{N-n}} \qquad Q.E.D.$$

Although $N$ is left out as a free parameter in the LB(N) algorithm, it need not be significantly larger than $n$. For example, if $N = n + 6$, then $P_{err} < 1.61 * 10^{-28}$. As argued in [14, § 4.5.4], the probability of an error in a computer circuit due to hardware malfunction or cosmic radiation is larger than the above bound.

***Theorem 2.*** *Assume a satisfiability problem with m clauses of n Boolean variables. Furthermore assume that we implement the LB(N) algorithm using K-input quantum gates ($K \geq 2$) that operate in time $t_K$; that the copy function $\mathbf{C}(\ )$ takes time $t_C$; and that the time to create a uniformly distributed qubit is $t_q$. Then the time complexity of LB(N) is*

$$T_{LB(N)} = O(t_q n + t_K m \log_K n + (t_K + t_C) N)$$

*which reduces to $O(t_q n + t_K m \log_K n + t_C n)$, when (as expected) $N = O(n)$.*

***Proof***

Let $T_i$ be the time consumed by the $i$-th step of the algorithm ($i = 1, 2, 3, 4, 5$). Then it is:

$T_1 = O(t_q n)$     (to create the $n$ qubits)
$T_2 = O(t_K m \log_K n)$     (since we can implement an ‖-gate of $O(n)$ inputs with $\log_K n$ layers of $K$-input ‖-gates)
$T_3 = O(t_K \log_K m)$     (since we can implement an &-gate of $m$ inputs with $\log_K m$ layers of $K$-input &-gates)
$T_4 = O((t_C + t_K) N)$     (to copy and boost $N$ times sequentially)
$T_5 = O(1)$     (to look up a given qubit)

Adding the above times together…     *Q.E.D.*

Setting $t = \max\{t_q, t_K, t_C\}$, the time complexity of LB(N) becomes $T_{LB(N)} = O(t(N + m\log_K n))$. Since $N$ should be $O(n)$ and $m$ would usually be much larger than $n$, the complexity reduces to $T_{LB(N)} = O(tm\log_K n)$. And because $t$ would be a constant independent of $n$ and $m$, the complexity of the LB(N) algorithm is $T_{LB(N)} = O(m\log_K n)$.

Given that satisfiability is NP complete, and that, depending on the nature of the cloning oracle, either binary random variables or qubits could be used, the above theorems establish that $NP \subseteq BPP^C \subseteq BQP^C$.

## 5. Approximate cloning

Unfortunately, as shown in [19], it is not possible to create a cloning function like **C**( ) using linear evolution. Although, there has been so far no such proof for non-linear systems, it is not very reasonable to believe--given the already proven fact that cloning leads to solution of NP complete problems-- that non-linear systems capable of exact cloning would be realizable (in the near future).

The above restrictions lead us to consider approximate cloning: a function AC( ) that can generate an approximate copy of any binary random variable or qubit *X*. That approximate copy should have the property that $|P(X = 0) - P(AC(X) = 0)| \leq \varepsilon$. In that case, we would say that AC(*X*) is an *ε-approximation* of *X* and that $\varepsilon$ is the *approximation degree* of AC(*X*). ($\varepsilon$ would normally be a small nonnegative number.)

Consequently, we can define the approximate logical boosting algorithm (ALB(N)) to be the LB(N) algorithm where **C**( ) has been replaced by AC( ). We shall now prove that if $\varepsilon$ is exponentially small (wrt *n*) then the error of ALB(N) is trivial and negligible.

**Theorem 3.** *The ALB(N) algorithm asserts that a given satisfiability problem S is satisfiable or unsatisfiable with a probability of error $P_{err}$, bounded:*
$$P_{err} < \max\left\{2^{-7(N-n)+34}, (2^N - 1)\varepsilon\right\}$$
*where $n \geq 7$ is the number of Boolean variables, N is the boosting level chosen in the ALB algorithm, and $\varepsilon \leq 2^{-n-1}$ is the approximation degree of AC(X).*
*Proof*
In order to prove this theorem, we must first prove the following two lemmas:
**Lemma 1.** *If $\{d_k\}$ is a sequence of real numbers satisfying the property that $d_0 \leq 1 - 2^{-n}$ and $d_{k+1} \leq d_k(d_k + \varepsilon)$, where $\varepsilon \leq 2^{-n-1}$, $n \geq 7$, then it is $d_k < 2^{-7(k-n)+34}$.*
*Proof*
Apparently, $d_{k+1} \leq d_k \leq 1 - 2\varepsilon$ (the proof, by induction on k, is trivial). Therefore, $\{d_k\}$ is a decreasing sequence.
Now we shall prove that $d_{n+2} < \tfrac{2}{3}$ assuming that $d_{n+2} \geq \tfrac{2}{3}$ and deriving a logical contradiction. Thus, if $d_{n+2} \geq \tfrac{2}{3}$ then, for every $k \leq n+2$, it would be $d_k \geq \tfrac{2}{3}$ and:

$$\begin{aligned}
d_k &\leq d_{k-1}^2\left(1+\tfrac{\varepsilon}{d_{k-1}}\right) & \text{(by defintion of } \{d_k\}) \\
&\leq d_{k-1}^2\left(1+\tfrac{3\varepsilon}{2}\right) & (\text{since } d_{k-1} \geq \tfrac{2}{3}) \\
&\leq d_0^{2^k}\left(1+\tfrac{3\varepsilon}{2}\right)^{1+2+\ldots 2^{k-1}} & \text{(by induction on } k) \\
&< \left\{d_0\left(1+\tfrac{3\varepsilon}{2}\right)\right\}^{2^k} & (\text{since } 1+2+\ldots+2^{k-1} < 2^k) \\
&\leq \left\{(1-2^{-n})\left(1+\tfrac{3*2^{-n-1}}{2}\right)\right\}^{2^k} & (\text{since } d_0 \leq 1-2^{-n} \text{ and } \varepsilon \geq 2^{-n-1}) \\
&\leq \left\{1-2^{-n-2}\right\}^{2^k} \\
&< \left(\tfrac{1}{e}\right)^{2^{k-(n+2)}} & (\text{since } (1-\tfrac{1}{L})^L < \tfrac{1}{e})
\end{aligned}$$

The latter inequality, for $k=n+2$, leads us to the desired contradiction.

Now, the proved statement, $d_{n+2} < \tfrac{2}{3}$, by repeated applications of the defining inequality of $\{d_k\}$ and the fact that $\varepsilon \leq 2^{-n-1}$, $n \geq 7$, leads to the following bounds: $d_{n+3} < \tfrac{1}{2} - 4\varepsilon$, $d_{n+4} < \tfrac{1}{4} - 3\varepsilon$, $d_{n+5} < \tfrac{1}{16} - \varepsilon$, $d_{n+6} < \tfrac{1}{256}$, $d_{n+6}+\varepsilon < \tfrac{1}{128}$; the last two now being used to complete this proof:

$$d_k \leq d_{k-1}(d_{k-1}+\varepsilon) < \tfrac{d_{k-1}}{128} < \ldots \leq \tfrac{d_{n+6}}{128^{k-(n+6)}} < \tfrac{1}{128^{k-(n+6)}*256} = \tfrac{1}{2^{7(k-n)-34}} \qquad Q.E.D.$$

**Lemma 2.** *If $\{d_k\}$ is a sequence of real numbers satisfying the property that $d_0 = 1$ and $d_{k+1} \geq d_k(d_k - \varepsilon)$, then it is $d_k \geq 1-(2^k-1)\varepsilon$ (equality holds only for $k=1$).*

*Proof*

We shall prove the lemma by induction on $k$:

$k=1$: $d_1 \geq d_0(d_0 - \varepsilon) = 1-\varepsilon$ (equality holds indeed).

$k=2$: $d_2 \geq d_1(d_1 - \varepsilon) = (1-\varepsilon)(1-2\varepsilon) > 1-3\varepsilon$ (equality holds indeed).

Induction step: Assume the mentioned inequality for $k$, and prove it for $k+1$:

$$\begin{aligned}
d_{k+1} &\geq d_k(d_k - \varepsilon) & \text{(by defintion of } \{d_k\}) \\
&\geq \{1-(2^k-1)\varepsilon\}\{1-2^k\varepsilon\} & \text{(by induction hypothesis)} \\
&= 1-(2^k+2^k-1)\varepsilon + 2^k(2^k-1)\varepsilon^2 \\
&> 1-(2^{k+1}-1)\varepsilon
\end{aligned}$$

Q.E.D.

Now, it is

$$\begin{aligned}
P_{err} &= P(D_N = 0/k_S > 0)P(k_S > 0) + P(D_N = 1/k_S = 0)P(k_S = 0) \\
&= P_{err,0}(1-p_S) + P_{err,1}p_S \\
&\leq \max\{P_{err,0}, P_{err,1}\}
\end{aligned}$$

where, $P_{err,0} = P(D_N = 0/k_S > 0)$ is the probability of error when $S$ is satisfiable, $P_{err,1} = P(D_N = 1/k_S = 0)$ is the probability of error when $S$ is unsatisfiable, and $p_S = P(k_S = 0)$ is the a-priori probability that problem $S$ is unsatisfiable. Furthermore, defining $\{d_v\}$ as in theorem 1 ($d_v = P(D_v = 0)$, $v = 0, 1, 2,\ldots, N$), we have

$$d_{k+1} = P(D_k \| AC(D_k) = 0) = P(D_k = 0)P(AC(D_k) = 0) = d_k(d_k + \varepsilon_k)$$

where $-\varepsilon \leq \varepsilon_k \leq \varepsilon$ is the approximation error of $AC(D_k)$ (at $k$-th step). As a result

$$d_k(d_k - \varepsilon) \leq d_{k+1} \leq d_k(d_k + \varepsilon)$$

and thus,

$$\begin{aligned}
P_{err,0} &= d_N\big|_{d_0 \leq 1-2^{-n}} &\leq 2^{-7(N-n)+34} & \text{(by lemma 1)} \\
P_{err,1} &= 1-d_N\big|_{d_0=1} &\leq (2^N-1)\varepsilon & \text{(by lemma 2)}
\end{aligned}$$

Finally, we can substitute the above in the definition of $P_{err}$ \hfill Q.E.D.

A consequence of the above theorem is that, by selecting $N = n+12$, $\varepsilon \leq 2^{-n-62}$, we get a trivial error: $P_{err} < 2^{-50}$ [14, § 4.5.4]. Furthermore, as it was shown in the above proof, $d_{n+4} < \frac{1}{4} - 3\varepsilon$. Therefore, selecting $N = n+4$ and $\varepsilon = 2^{-n-6}$, we get a probability of error less than ¼; which means that the ALB($n$+4) algorithm is proof of membership of satisfiability in BQP$^{AC}$ (or BQP$^{AC}$, depending on the cloning oracle).

However, the fact that $\varepsilon$ is exponentially small (with respect to $n$) is a very hard constraint. At the time of this writing and to the best of the authors' knowledge, it is not known whether such high precision approximate cloning is possible. When unitary evolution is used, provably optimal approximate cloning, starting from N qubits and generating M clones at the end, results in fidelity $\frac{M(N+1)+N}{M(N+2)}$, or equivalently precision $\varepsilon = \frac{M-N}{M(N+2)}$ [13]. That is very low when compared to $\varepsilon = 2^{-n-6}$, which is required by the ALB($n$+4) algorithm. The use of non-linearity may improve upon the situation (since cloning is a non-linear operation) but it remains an open research problem the maximum possible precision and its time complexity. The fact that cloning leads to solution of the NP-complete problems only stresses its importance.

On the other hand, if we consider low precision cloning, i.e. $\varepsilon = O(1/p(n))$, where $p(n)$ is a polynomial of $n$, then, since $d_0$ can be as large as 1-$2^{-n}$, there are no bounds on the probability of error (other than the trivial ones). However, it is possible, in that case, to treat the algorithm in a *probably approximately correct* fashion; that is to question whether $P(P_{err} > \delta) < \gamma$, for some small $\delta$, $\gamma$. However, that analysis is beyond the scope of this paper and may be presented elsewhere.

## 6. Non-cloning logical boosting (fixed point)

The uncertainty about high precision approximate cloning leads to other considerations: perhaps we can transform $D_k$ to $D_{k+1}$ ($k = 0,1,…, N$-1) using unitary evolution and logic.

Let $\mathbb{R}_+$ is the set of nonnegative real numbers, $\mathbb{C}$ be the set of complex numbers, $t$ indicate matrix and vector transpose, and $|z|$ denote the magnitude of $z \in \mathbb{C}$. Each qubit $Q$, represented by a vector of two complex numbers $[q_0, q_1]^t$ ($|q_0| = P(Q = 0)$, $|q_1| = P(Q = 1)$, $|q_0|^2+|q_1|^2 = 1$), corresponds to a point (or vector) in $\mathbb{C}^2$. Furthermore, if $x = [x_0,…,x_{n-1}]^t \in \mathbb{C}^n$, we define

$$|x| = \sqrt{|x_0|^2 + … + |x_{n-1}|^2} \quad \text{and} \quad \langle x \rangle = [|x_0|,…,|x_{n-1}|]^t$$

Finally, a function $f : \mathbb{C}^n \to \mathbb{C}^n$ has a *magnitude fixed point* (mfp) $x \in \mathbb{R}_+^n$ if and only if $\forall y \in \mathbb{C}^n (\langle y \rangle = x \Rightarrow \langle f(y) \rangle = x)$. (Apparently, if $f$ has an mfp $x$, then $\langle f(x) \rangle = x$.)

The *Unitary Boosting Algorithm* (UB(N)) is like LB(N) except that, instead of using a qubit copy/cloning function, it transforms $D_k$ to $D_{k+1}$ using unitary matrix $U^{(k)}$ and logical circuit $L^{(k)}$, i.e. $D_{k+1} = L^{(k)}(U^{(k)} \cdot D_k P^{(k)})$, where $P^{(k)}$ is a set of normalized internal parameters ( $P^{(k)} \in \mathbb{C}^{2^p}$, $|P^{(k)}| = 1$, $U^{(k)} \in \mathbb{C}^{2^{p+1} \times 2^{p+1}}$ ).

Note that, if we desire that UB(N) never errs when it asserts that $S$ is satisfiable, it must be $d_N = 1$ when $d_0 = 1$; or, equivalently $\langle D_N \rangle = [1, 0]^t$ when $\langle D_0 \rangle = [1, 0]^t$ -- which, in turn, means that the overall transformation of $D_0$ to $D_N$ has an mfp at $[1, 0]^t$. This later property is implied by the condition that the transformation of $D_k$ to $D_{k+1}$ (for all $k$) has an mfp at $[1, 0]^t$. However, as the following theorem shows, it is impossible to reduce $d_k$ in that case.

**Theorem 4.** If $D_{k+1} = L(U \cdot D_k H)$, where $D_k$ and $D_{k+1}$ are qubits ($D_k, D_{k+1} \in \mathbb{C}^2$), $H$ is a unit-length vector of $2^h$ complex numbers ($H \in \mathbb{C}^{2^h}$ is the set of internal of hidden parameters corresponding to $h$ qubits), $U$ is a $2^{h+1} \times 2^{h+1}$ unitary matrix ($U \in \mathbb{C}^{2^{h+1} \times 2^{h+1}}$), $L$ is an arbitrary logical circuit of $h+1$ inputs, and the overall transformation has an mfp at $[1,0]^t$, then $d_{k+1} \geq d_k$.

**Proof**

If $W$ is a matrix, let $W_{i,j}$ denote the $i$-th row and $j$-th column element of $W$, $W_j$ denote the $j$-th column of $W$, and thus $(W^t)_i$ denote the vector correspoding to the i-th row of $W$. Similarly, if $V$ is a vector, let $V_i$ be the i-th element of $V$.

Now let $\{R_0, \ldots, R_{2^h-1}\}$ be an orthonormal base of $\mathbb{C}^{2^h}$, where $R_0 = \overline{H}$, and form the matrix

$$X^t = \begin{bmatrix} R_0 & \cdots & R_{2^h-1} & 0 & \cdots & 0 \\ 0 & \cdots & 0 & R_0 & \cdots & R_{2^h-1} \end{bmatrix} \in \mathbb{C}^{2^{h+1} \times 2^{h+1}}$$

(Apparently, $\{(X^t)_0, \ldots, (X^t)_{2^h-1}\}$ is an orthonormal base of $\mathbb{C}^{2^{h+1}}$.)

Now define $A = UX^{-1}$, or, equivalently, $U = AX$, and $D_k = [a_k, b_k]^t$. As a result, $A$ is unitary and

$$UD_k H = AX[a_k P, b_k H]^t = A[a_k, 0, \ldots, 0, b_k, 0, \ldots, 0]^t = a_k A_0 + b_k A_{2^h}$$

and if we indicate by $T_L$ and $F_L$ the sets of integers (in the interval $[0, 2^{h+1}-1]$) whose binary representation corresponds to logical assignments that make $L$ to yield true or false, respectively, then it would be,

$$\begin{aligned}
d_{k+1} &= P(L(UD_k H) = 0) && \text{(by the definition of } d_{k+1}, D_{k+1}) \\
&= \sum_{i \in F_L} |(a_k A_0 + b_k A_{2^h})_i|^2 && \text{(by the prev. result about } UD_{k+1}H) \\
&= |a_k|^2 \sum_{i \in F_L} |A_{i,0}|^2 + |b_k|^2 \sum_{i \in F_L} |A_{i,2^h}|^2 + Y + \overline{Y} && \text{(where } Y = \overline{a}_k b_k \sum_{i \in F_L} \overline{A_{i,0}} A_{i,2^h}) \\
&= |a_k|^2 + |b_k|^2 \sum_{i \in F_L} |A_{i,2^h}|^2 + Y + \overline{Y} && \text{(assuming } \sum_{i \in F_L} |A_{i,0}|^2 = 1) \\
&= |a_k|^2 + |b_k|^2 \sum_{i \in F_L} |A_{i,2^h}|^2 && \text{(assuming } Y = 0) \\
&\geq d_k && \text{(since } d_k = |a_k|^2)
\end{aligned}$$

To complete the proof, we must prove the premises $\sum_{i \in F_L} |A_{i,0}|^2 = 1$ and $Y=0$ (both of which result from the fact that our $L$-$U$ transformation has an mfp at $[1, 0]^t$). Indeed, when $D_k = [1, 0]^t$ then $1 = d_{k+1} = L(U[1,0]^t H) = \sum_{i \in F_L} |A_{i,0}|^2$. Finally, the fact that $U$ (and thus $A$) is unitary implies that:

1. $|A_0|^2 = 1 \Rightarrow \sum_{i \in F_L} |A_{i,0}|^2 + \sum_{i \in T_L} |A_{i,0}|^2 = 1 \Rightarrow \sum_{i \in T_L} |A_{i,0}|^2 = 0 \Rightarrow \forall i \in T_L (A_{i,0} = 0)$

2. $\overline{A_0} A_{2^h} = 0 \Rightarrow \sum_{i \in F_L} \overline{A_{i,0}} A_{i,2^h} + \sum_{i \in T_L} \overline{A_{i,0}} A_{i,2^h} = 0 \Rightarrow \sum_{i \in F_L} \overline{A_{i,0}} A_{i,2^h} = 0 \Rightarrow Y = 0$

Q.E.D.

The above theorem demonstrates that no combination of such unitary transformations and logic can solve the NP complete problems (probabilistically).

Whether there is a sequence of unitary matrices and logical circuits which constitute transformations with no mfp at $[1, 0]^t$ and result in an algorithm with small probability of error is still an open research question.

## 7. Conclusion

NP is a very interesting class of problems of extensive and tremendous practical significance. NP complete problems (the hardest in NP) seem, at first sight, to be irrelevant to cloning. However, their direct relationship to cloning and therefore the inherent difficulty of the latter has now been established.

Exact cloning is perhaps impossible since it is an operation of infinite precision. Approximate cloning is possible but being a non-linear operation, it is perhaps best achieved through non-linear operations. Linear (and unitary) approaches offer very limited precision, which does not allow for the solution of large NP complete problems. It is therefore important to try to develop cloning methods that would employ non-linearity to gain precision.

Finally, it is not clear whether it would be possible to replace cloning in the ALB(N) algorithm by a sequence of (unitary and non-linear) transformations that would result in a final binary random variable or qubit with trivial (or small) probability of error. A straightforward attempt was proven fruitless, but there is no evidence in either direction about more elaborate schemes. The matter is an open research question.

**Acknowledgements**

This work was partially supported by the European Union RTN grants HPRN-CT-2000-00122 and -00131.

**References**

1. P. Benioff, "Quantum mechanical Hamiltonian models of Turing machines," *Journal of Statistical Physics,* vol. 29, pp. 515-546, 1982.
2. P. Benioff, "Quantum mechanical Hamiltonian models of Turing machines that dissipate no energy," *Physics Review Letters*, vol. 48, pp. 1581-1585, 1982.
3. C. H. Bennett, E. Bernstein, G. Brassard, and U. Vazirani, "Strengths and weaknesses of quantum computing," *SIAM Journal on Computing,* vol. 26, no. 5, pp. 1510-1523, 1997.
4. E. Bernstein and U. Vazirani, "Quantum complexity theory," *Proceedings of 25th ACM Symposium on Theory of Computing,* pp. 11-20, 1993.
5. A. Berthiaume and C. Brassard, "The quantum challenge to structural complexity theroy," *Proceeding of 7th IEEE Conference on Structure in Complexity Theory,"* pp. 132-137, 1992.

6. A. Berthiaume and C. Brassard, "Oracle quantum computing," *Journal of Modern Optics,* 41, pp. 2521-2535, 1994.
7. M. Boyer, C. Brassard, P. Hoyer, and A. Tapp, "Tight bounds on quantum searching," *Proceedings of 4th Workshop on Physics and Computation,* Boston, pp.36-43, 1996. (Available online: `http://interjournal.org`.)
8. D. Deutsch, "Quantum theory, the Church-Turing principle and the universal quantum computer," *Proceedings of Royal Society, London,* ser. A, vol. 425, pp. 73-90, 1989.
9. D. Deutsch, "Quantum computational networks," *Proceedings of Royal Society, London,* ser. A, vol. 439, pp. 553-558, 1992.
10. D. Deutsch and R. Jozsa, "Rapid solution of problems by quantum computation," *Proceedings of Royal Society, London,* ser. A, vol. 439, pp. 553-558, 1992.
11. R. Feynman, "Simulating physics with computers," *International Journal of Theoretical Physics,* vol. 21, No. 6/7, pp. 467-488, 1982.
12. R. Feynman, "Quantum mechanical computers," *Foundations of Physics,* vol. 16, pp. 507-531, 1986.
13. N. Gisin and S. Massar, "Optimal quantum cloning machines," Physical Review Letters, 79, pp. 2153-2156, 1997.
14. D. Knuth, *The Art of Computer Programming, Vol 2: Seminumerical Algorithms.* Addison-Wesley, Reading, Mass., 1981.
15. S. Lomonaco, Jr., "A quick glance at Quantum Cryptography," http://xxx.lanl.gov/abs/quant-ph/9811056.
16. C. H. Papadimitriou, *"Computattional Complexity,"* Addison-Wesley, 1994.
17. P. Shor, "Algorithms for quantum computation: Discrete logarithms and factoring," *Proceedings of 35th IEEE Symposium on Foundations of Computer Science,* pp. 124-134, 1994.
18. D. Simon, "On the power of quantum computation," *Proceedings of 35th Symposium on Foundations of Computers Science,* pp. 116-123, IEEE Computer Society Press, 1994.
19. W. K. Wooters and W. H. Zurek, "A single quantum cannot be cloned," *Nature,* 299, pp. 802-803, 1982.
20. A. Yao, "Quantum circuit complexity," *Proceedings of 34th IEEE Symposium on Foundations of Computer Science,* pp. 352-361, 1993.